\documentclass[11pt]{article}
\usepackage{amsmath,amssymb}
\usepackage{lmodern,eucal}
\usepackage[utf8]{inputenc}
\usepackage[T1]{fontenc}
\usepackage{graphicx}
\usepackage[footskip=0.5cm,margin=1.5cm]{geometry}
\usepackage[numbers,compress,super]{natbib}
\usepackage[font=small,labelfont=bf]{caption}
\DeclareCaptionLabelSeparator{line}{.}
\usepackage[labelsep=line]{caption}
\usepackage{placeins}
\usepackage{lineno}

\usepackage[numbers,compress]{natbib}
\usepackage{hyperref}
\usepackage{xcolor}
\definecolor{blue}{HTML}{306EFF}
\definecolor{darkgray}{HTML}{242220} 
\hypersetup{colorlinks=true, urlcolor=blue, citecolor=blue, linkcolor=blue}
\color{darkgray}

\usepackage{subcaption}
\usepackage{titlesec}
\titleformat{\section}[block]{\bfseries}{\thesection.}{0.3em}{}
\titleformat{name=\section,numberless}[block]{\large\bfseries}{}{0em}{}
\titleformat{\subsection}[block]{\itshape\centering}{\thesubsection.}{0.3em} {}

\usepackage{footmisc}

\usepackage{soul}
\setul{}{1.5pt}
\setstcolor{red}

\renewcommand{\l}[0]{\left}
\renewcommand{\r}[0]{\right}
\let\f=\frac

\renewcommand{\v}[1]{\mathbf{#1}}
\newcommand{\uv}[1]{\mathbf{\hat{#1}}}
\newcommand{\abs}[1]{\l| #1 \r|}

\newcommand{\avg}[1]{\langle #1 \rangle}
\newcommand{\Avg}[1]{\l< #1 \r>}

\newcommand{\Ang}[0]{\, \mathring{\mathrm{A}}}
\newcommand{\kB}[0]{k_\t{B}}
\newcommand{\K}[0]{\, \t{K}}

\renewcommand{\t}[1]{\text{#1}}

\begin{document}
  % Fix citation style.
  \makeatletter
  \def\bstctlcite{\@ifnextchar[{\@bstctlcite}{\@bstctlcite[@auxout]}} \def\@bstctlcite[#1]#2{\@bsphack\@for\@citeb:=#2\do{\edef\@citeb{\expandafter\@firstofone\@citeb} \if@filesw\immediate\write\csname #1\endcsname{\string\citation{\@citeb}}\fi} \@esphack}
  \makeatother
  \bstctlcite{fix_citation_style}

  % Title page.
  \nolinenumbers
  \begin{center}
    \textbf{\large Quantum effects on dislocation motion from Ring-Polymer Molecular Dynamics} \\
    \vspace{1.5cm}
    {\large Rodrigo Freitas\footnote{Corresponding author (email: freitas@stanford.edu).}\footnote{Current affiliation: Department of Materials Science and Engineering, Stanford University, Stanford, CA, 94305, USA}} \\
    \textit{\small Department of Materials Science and Engineering, University of California, Berkeley, CA, 94720, USA} \\
    \textit{\small Lawrence Livermore National Laboratory, CA, 94550, USA} \\
    \vspace{0.8cm}
    {\large Mark Asta} \\
    \textit{\small Department of Materials Science and Engineering, University of California, Berkeley, CA, 94720, USA} \\
    \textit{\small Materials Sciences Division, Lawrence Berkeley National Laboratory, Berkeley, CA, 94720, USA} \\
    \vspace{0.8cm}
    {\large Vasily V. Bulatov} \\
    \textit{\small Lawrence Livermore National Laboratory, Livermore, CA, 94550, USA} \\
    \vspace{0.2cm}
    \small{(Dated: \today)}
  \end{center}
  \vspace{1.5cm}

  \begin{abstract}
    Quantum motion of atoms known as zero-point vibration was recently proposed to explain a long-standing discrepancy between theoretically computed and experimentally measured low-temperature plastic strength of iron and possibly other metals with high atomic masses. This finding challenges the traditional notion that quantum motion of atoms is relatively unimportant in solids comprised of heavy atoms. Here we report quantum dynamic simulations of quantum effects on dislocation motion within the exact formalism of Ring-Polymer Molecular Dynamics (RPMD). To extend the reach of quantum atomistic simulations to length and time scales relevant for extended defects in materials, we implemented RPMD in the open-source code LAMMPS thus making the RPMD method widely available to the community. We use our RPMD/LAMMPS approach for direct calculations of dislocation mobility and its effects on the yield strength of $\alpha$-iron. Our simulation results establish that quantum effects are noticeable at temperatures below $50\K$ but account for only a modest ($\approx 13\%$ at $T = 0\K$) overall reduction in the Peierls barrier, at variance with the factor of two reduction predicted earlier based on the more approximate framework of harmonic transition state theory. Our results confirm that zero-point vibrations provide ample additional agitation for atomic motion that increases with decreasing temperature, however its enhancing effect on dislocation mobility is largely offset by an increase in the effective atom size, an effect known as quantum dispersion that has not been accounted for in the previous calculations.
  \end{abstract}
  \vspace{0.5cm}
  \newpage

  \noindent \textbf{\large Introduction}

    Quantum motion of atoms known as zero-point vibrations is recognized to be important at low temperatures in condensed matter systems comprised of light atoms or ions, affecting such properties and behaviors as proton-transfer reactions\cite{proton_1,proton_2}, vibrational spectra of water\cite{water_1,water_2} and ice\cite{ice_1,ice_2}, and mechanical properties of low temperature helium\cite{he_2,he_3,he_1}. Recently, quantum motion of atoms was proposed to explain a long-standing discrepancy between theoretically computed and experimentally measured low-temperature resistance (Peierls stress) to dislocation motion in iron and possibly other metals with high atomic masses\cite{qtst_neb_1,qtst_neb_2,qtst_neb_3}. 

    Experimental estimates for the Peierls stress $\tau_\t{P}$ in body-centered cubic (bcc) metals are obtained by relating this parameter to the low-temperature yield strength of the metal extrapolated to zero temperature. In $\alpha$-iron, the so-obtained experimental estimates for $\tau_\t{P}$ range between $0.35\,\t{GPa}$ and $0.45\,\t{GPa}$\cite{ps_exp_1,ps_exp_2,ps_exp_3,ps_exp_4} whereas atomistic calculations consistently predict $\tau_\t{P}$ to be much higher, ranging between $0.9\,\t{GPa}$ and $1.2\,\t{GPa}$ depending on the interatomic potential model used in the calculations\cite{ps_md_1,ps_md_2,qtst_neb_1}. At first perceived as a failure of interatomic potentials, subsequent electronic structure calculations confirmed the stubbornly high theoretical values of $\tau_\t{P}$ in $\alpha$-iron to be between $1.0\,\t{GPa}$ and $1.4\,\t{GPa}$\cite{ps_dft_1,ps_dft_2,ps_dft_3}. In ref.~\citenum{qtst_neb_1} the effect of zero-point vibrations was accounted for by computing an approximate quantum correction to the rate of dislocation motion predicted within the classical transition state theory (TST)\cite{qtst_neb_1}. The resulting correction reduced the Peierls stress from its fully classical value of 0.9 GPa by about $50\%$, to $\tau_\t{P}=0.45\,\t{GPa}$ thus bringing it in close agreement with the experimental estimates.

    Although appealing, the dramatic two-fold reduction in the Peierls stress was predicted from a harmonic approximation to the classical TST in which classical populations of vibrational modes were replaced with their quantum analogs. Given its potential importance, here we re-examine the effect of zero-point vibrations on dislocation motion in $\alpha$-iron using Ring-Polymer Molecular Dynamics\cite{rpmd_original,rpmd_review} simulations (RPMD). Based on Feynman's path integral formulation of quantum mechanics, RPMD is a method that fully accounts for all effects of quantum motion of atoms on equilibrium statistical properties. As detailed in fig.~S1, RPMD amounts to a simultaneous simulation of P replicas of the entire N-atom system in which each atom is represented by its P clones connected into a ``polymer ring'' (a closed Feynman path) by elastic springs. RPMD is asymptotically exact in the limit $P\rightarrow\infty$. To take full advantage of RPMD accuracy, in the following we make no assumptions on the character of dislocation motion and measure the effect of zero-point vibrations on the Peierls stress in the most direct manner possible, akin to the very method by which the Peierls stress is estimated in experiment. Using the same interatomic model of $\alpha$-iron as in ref.~\citenum{qtst_neb_1}, here we compute the resistance to dislocation motion twice -- first using fully classical Molecular Dynamics (MD) and then using RPMD simulations -- and compare the results in the limit of zero temperature to assess the differences.

  \vspace{0.5cm}
  \noindent \textbf{\large Results and Discussion}

    As a baseline for subsequent comparison, first we compute the Peierls stress using fully classical simulations. Defined as the minimal stress required to make a dislocation move in the limit of zero temperature, Peierls stress can be computed in a static simulation in which stress or strain is increased in increments, each increment followed by a full relaxation of atom positions.  The Peierls stress is taken to be just one increment below the stress at which the dislocation begins to move. Our so-computed ``static'' Peierls stress is $\tau_\t{P}=(0.98\pm0.01)\,\t{GPa}$ for a $\f{1}{2}\avg{111}$ screw dislocation moving on a $\{112\}$ plane in the so-called twinning direction\cite{twinning_direction}. This value is close to the ones previously reported for the same potential\cite{qtst_neb_1,qtst_neb_2} ($\approx\!0.91\,\t{GPa}$). For reasons explained in Supplementary Information 1, the RPMD method does not permit similar static calculations which prompted us to compute the same Peierls stress again but now using classical dynamic simulations, to facilitate subsequent direct comparisons with RPMD. Another reason for us to perform dynamic simulations was an alarming, even if unnoticed, discrepancy in the literature between ``static'' and ``dynamic'' predictions for the Peierls stress reported by different research groups for the same model potential of iron\cite{ps_md_2,ps_dft_2,ps_dynamic}.

\begin{figure}[!htb]
      \centering
      \includegraphics[width=0.60\textwidth]{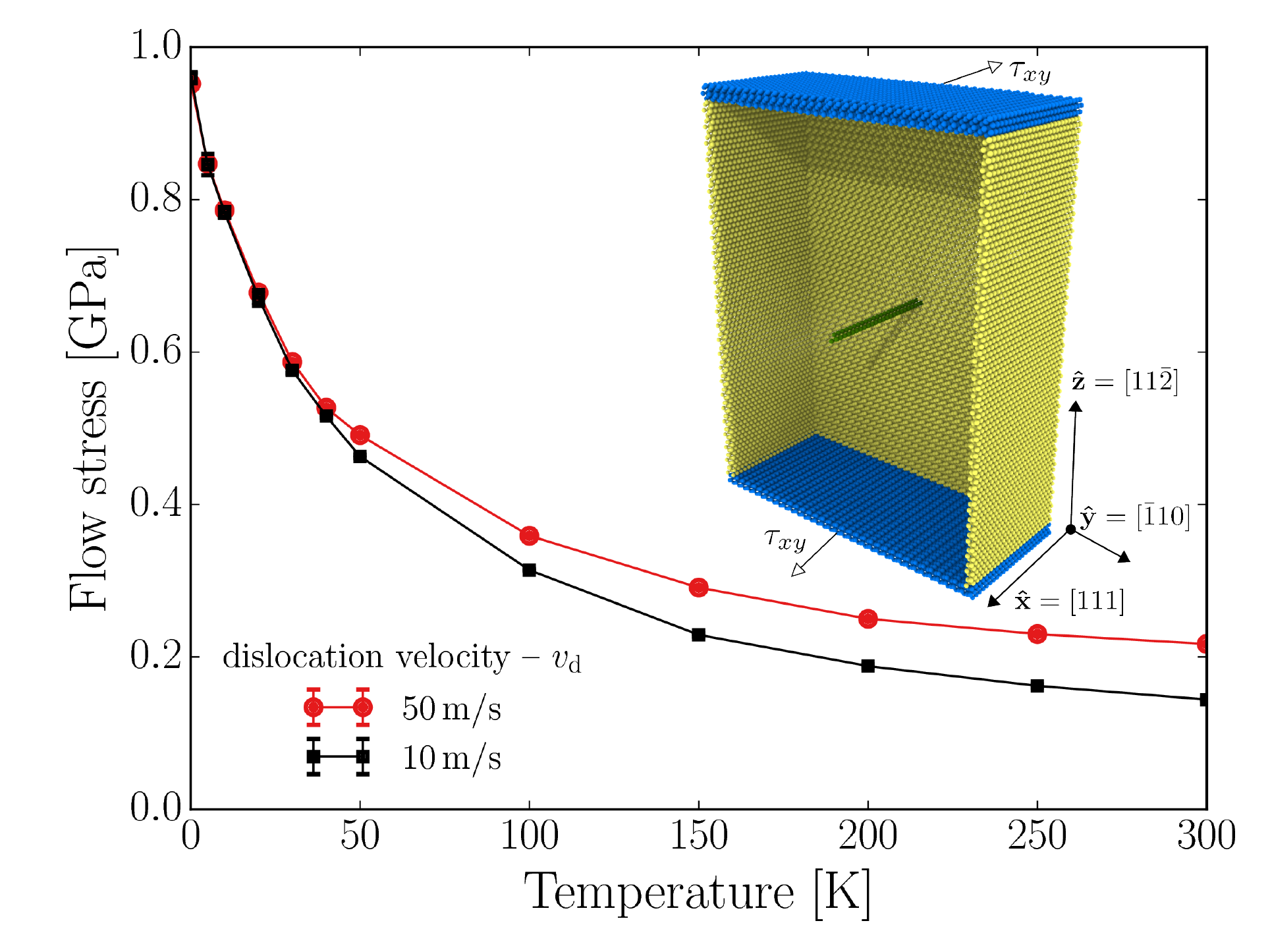}
      \caption{\label{fig:stress_vs_T_MD} Classical flow stress of a single screw dislocation in iron predicted in classical MD simulations as a function of temperature. The simulations were performed at two different dislocation velocities $v_\t{d}$: $50\,\t{m/s}$ (red symbols and red line) and $10\,\t{m/s}$ (black symbols and black line). The statistical error bars are smaller than the symbols. The inset depicts the simulation volume in which most atoms are deleted for clarity except for atoms closest to the core of a $\f{1}{2}\avg{111}$ screw dislocation near the center (green) and atoms at the boundaries of the simulation volume (blue and yellow). To persuade the dislocation to move along the horizontal $(11\bar{2})$ plane, the blue atoms in the top and in the bottom layers were moved along the $\uv{x}$ axis at constant and opposite velocities.}
    \end{figure}

    Shown in fig.~\ref{fig:stress_vs_T_MD} is the flow stress of a single screw dislocation extracted from our MD simulations, plotted as a function of simulation temperature. The flow stress is seen to decrease monotonically with increasing temperature which is typical of dislocations whose motion is temperature- and stress-activated: as the temperature is increased additional thermal energy becomes available helping the dislocation to overcome its motion barriers at an increased rate which requires lower stress to maintain the same dislocation velocity $v_\t{d}$. To make sure that our ``dynamic'' calculations of the Peierls stress are consistent and robust, we performed two series of MD simulations over the same range of temperatures in which the dislocation was forced to move at two different velocities $v_\t{d}$: $50\,\t{m/s}$ and $10\,\t{m/s}$ (see Methods \ref{sec:methods_MD}). As is seen in fig.~\ref{fig:stress_vs_T_MD}, the flow stress at any given temperature is clearly velocity-dependent, however the same Peierls stress is obtained by extrapolation to $T=0\K$: $\tau_\t{P}=(0.95\pm0.02)\,\t{GPa}$ at $50\,\t{m/s}$ and $\tau_\t{P}=(0.96\pm0.02)\,\t{GPa}$ at $10\,\t{m/s}$. Within small error bars, our two ``dynamic'' predictions agree with our own as well as previously published ``static'' predictions for the Peierls stress\cite{ps_md_2,ps_dft_2}.

\begin{figure}[!htb]
      \centering
      \includegraphics[width=0.60\textwidth]{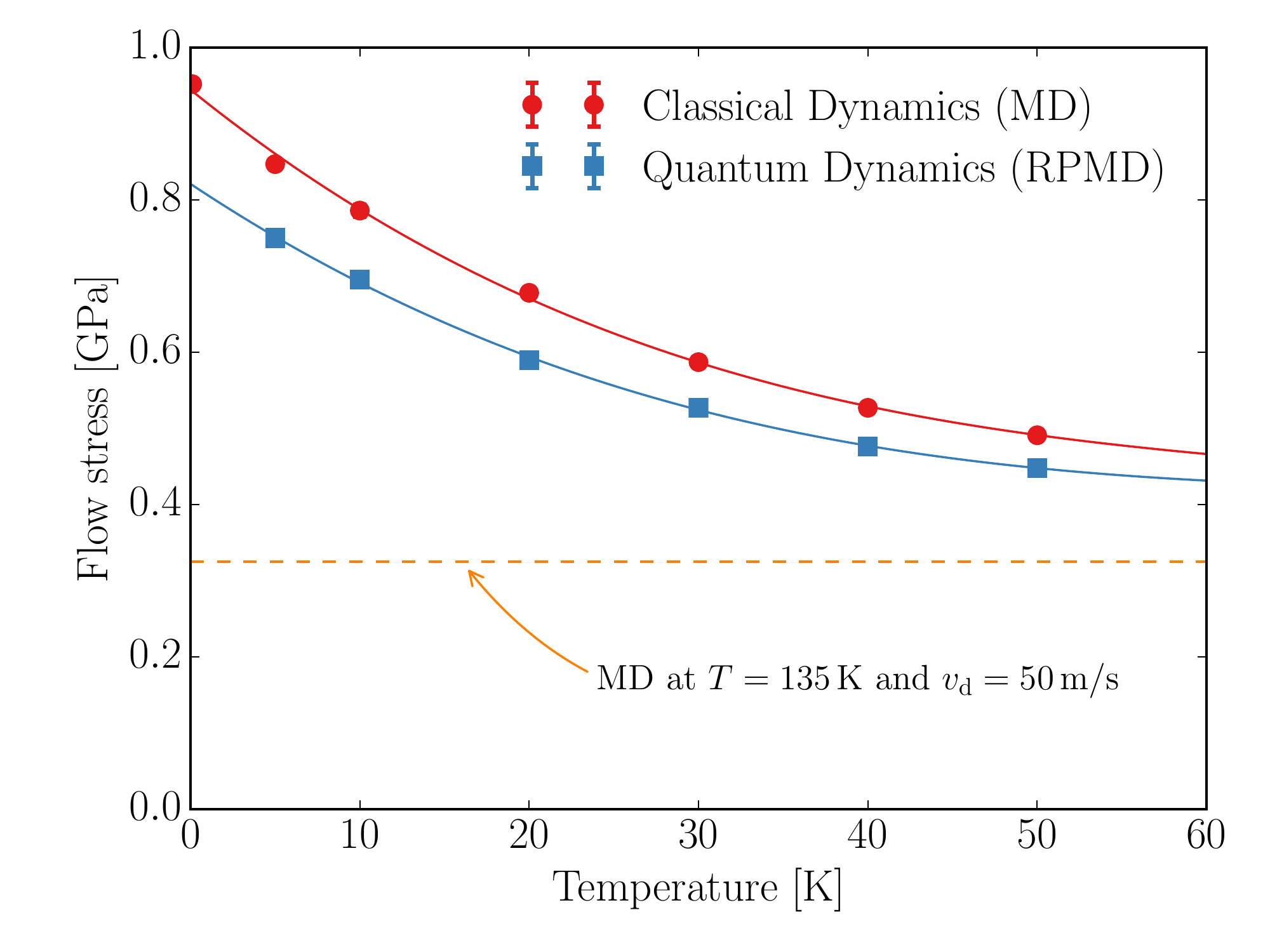}
      \caption{\label{fig:flow_stress} Temperature dependence of the flow stress of a single screw dislocation predicted in classical (red symbols and red line) and quantum (blue symbols and blue line) simulations. In both cases, the Peierls stress was obtained by a third-order polynomial (solid lines) extrapolation of the finite temperature flow stress to $T=0\K$. The orange dashed line is the flow stress as predicted from classical MD simulations at temperature $T = 135\K$ and dislocation velocity $v_\t{d} = 50 \, \t{m/s}$.}
    \end{figure}

    We now turn to RPMD simulations to compute the Peierls stress. Except for the method -- classical MD versus quantum RPMD -- our classical and quantum simulations were performed using the same interatomic potential and in otherwise identical simulation conditions (see Methods \ref{sec:methods_RPMD}). Figure \ref{fig:flow_stress} presents a comparison between MD and RPMD predictions of flow stress from $5\K$ to $50\K$. Following the same extrapolation procedure previously used to extract the Peierls stress from the classical MD data, the Peierls stress predicted in the quantum RPMD simulations is $13\%$ smaller than the classical MD result. In our Supplementary Information 2 we present substantial evidence that our implementation of the RPMD method is highly accurate, see figs.~S2, S3, and S4. Thus, we regard our prediction of only a modest reduction in the Peierls stress due to zero-point vibrations in $\alpha$-iron as more accurate than the dramatic $50\%$ reduction predicted earlier\cite{qtst_neb_1}.

    Although it may be difficult to separate factors that may have contributed to the substantial over-estimation of the effect of zero-point vibrations on the Peierls stress, several explicit or implicit assumptions used in ref.~\citenum{qtst_neb_1} to arrive at the dramatic prediction can be suspected. First, reliance on the fully classical TST in defining a critical pathway -- a minimum energy path (MEP) -- for quantum motion of atoms is inconsistent. Clearly, quantum-mechanical transition states relevant for dislocation motion must be defined and should be searched for in the extended space of closed Feynman paths\cite{qtst, rpmd_qtst_3}, i.e. in $3N\!P$-dimensional RPMD space. Another potential source of error is that Wigner correction (replacing the classical populations of vibrational states with the quantum ones) enhances the population of vibrational modes that accounts for the zero-point energy (ZPE) effects while ignoring other potentially important aspects of quantum motion. That something is amiss about limiting the effect of zero-point vibrations to just the Wigner correction is corroborated by the following qualitative argument.

\begin{figure}[!htb]
      \centering
      \includegraphics[width=0.60\textwidth]{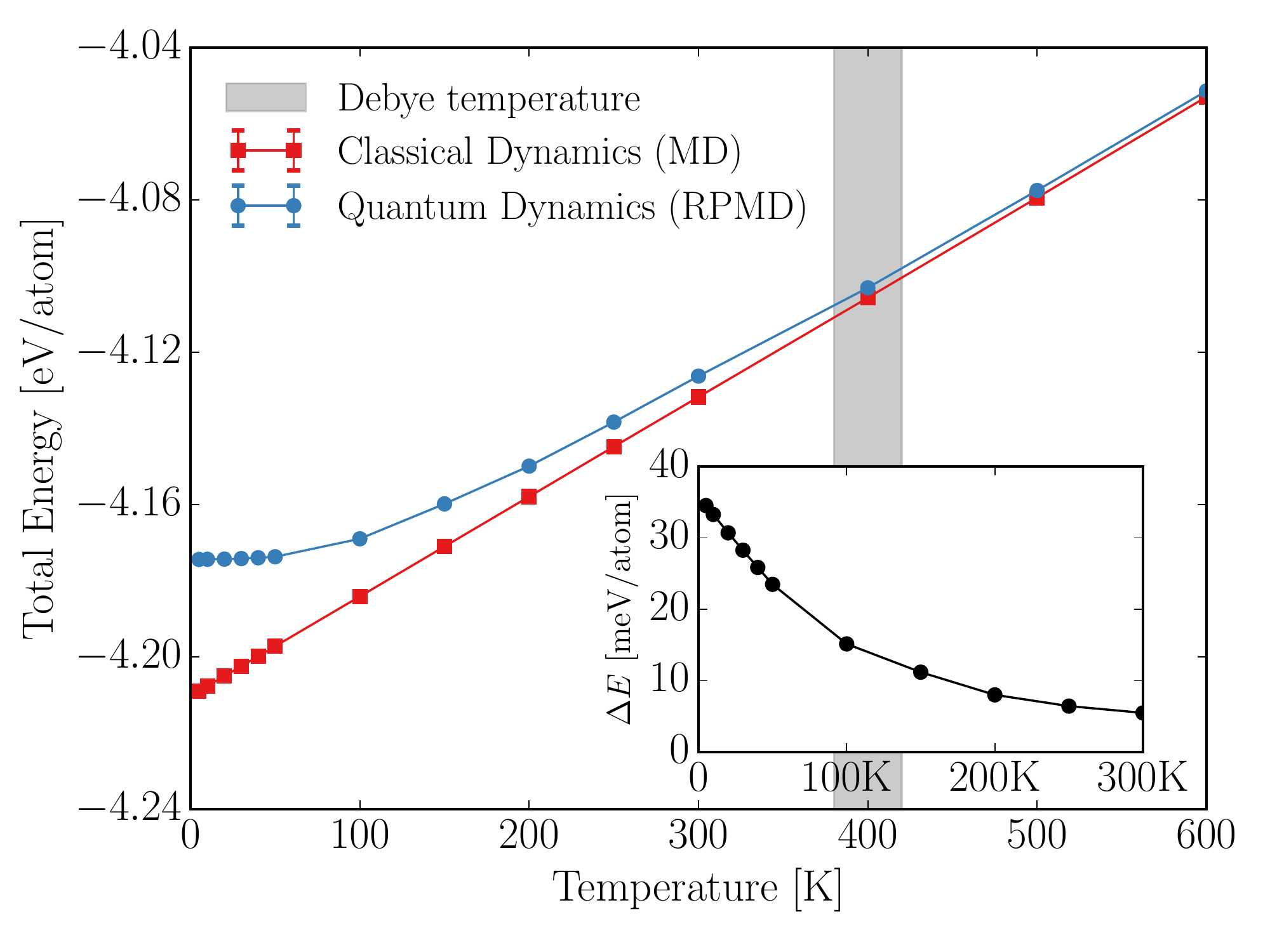}
      \caption{\label{fig:energy_vs_temperature} Temperature dependence of the per-atom energy in a perfect bcc lattice of $\alpha$-iron in the quantum (blue symbols and blue line) and the classical (red symbols and red line) systems. Shown in the inset is the excess energy of the quantum system equal to the ZPE at $T=0\K$. Shown as a gray vertical band is the Debye temperature of the model of $\alpha$-iron studied here, $T_\t{D}=(400\pm20)\K$.}
    \end{figure}

    As shown in fig.~\ref{fig:energy_vs_temperature}, the ZPE of a perfect crystal computed with the same interatomic potential model of iron (see Methods \ref{sec:methods_zpe_rg}) is $35\,\t{meV/atom}$. To approximately account for added agitation supplied in the form of zero-point vibrations, let us partition the ZPE equally between the potential and the kinetic energy per atom in the classical system, resulting in a classical temperature of $135\K$. By reference to our classical MD results for the flow stress, fig.~\ref{fig:stress_vs_T_MD}, this extra temperature would reduce the classical Peierls stress to $0.35\,\t{GPa}$, close to $0.45\,\t{GPa}$ reported in ref.~\citenum{qtst_neb_1}. That our explicit RPMD simulations predict a much more modest reduction suggests that some other effect(s) originating in quantum motion of atoms counterbalances this added agitation. We propose that such a contribution comes from quantum dispersion of atoms, i.e., finite size of a quantum particle compared to a point-particle in classical mechanics.

\begin{figure}[!htb]
      \centering
      \includegraphics[width=0.60\textwidth]{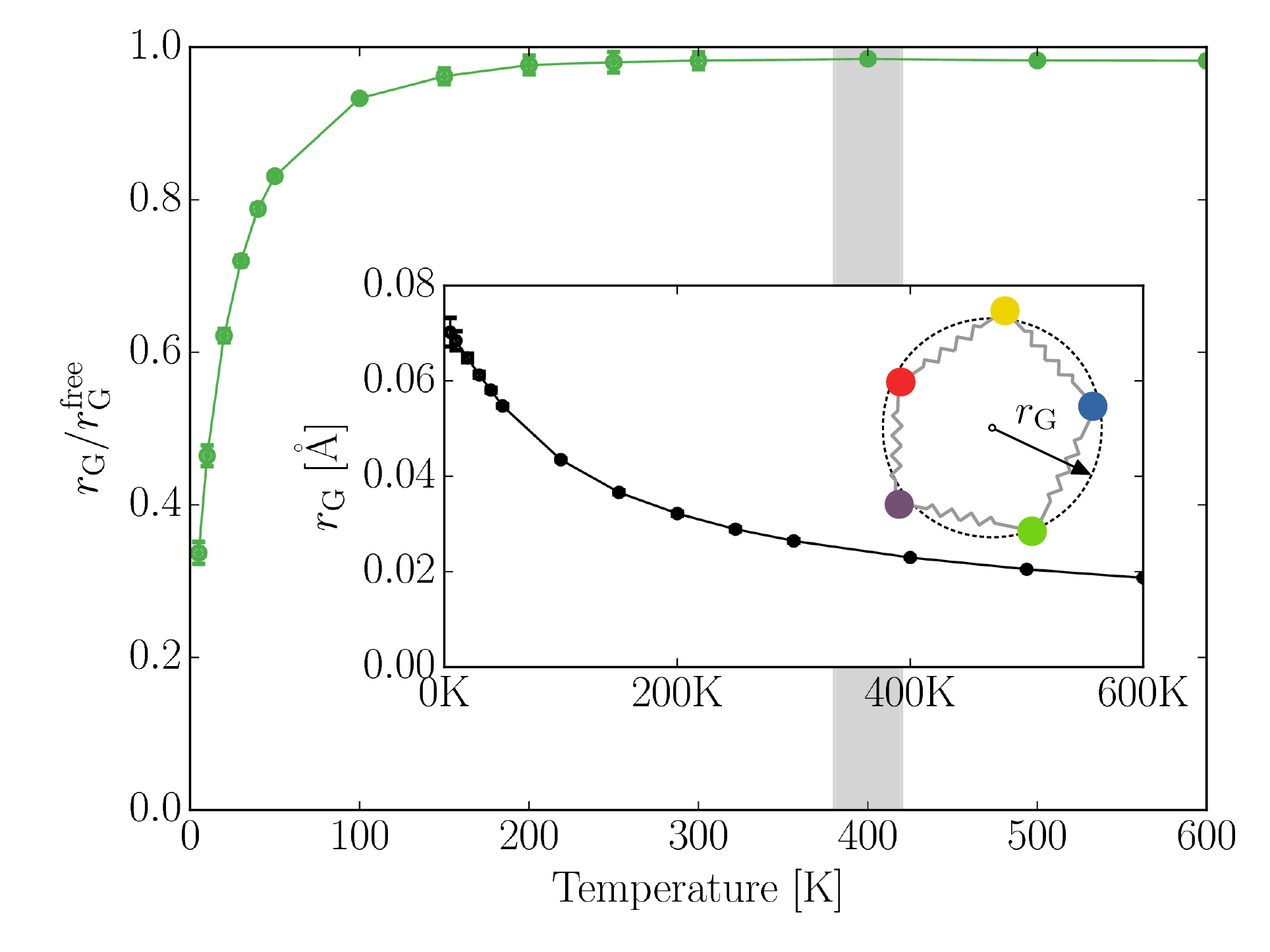}
      \caption{\label{fig:radius_of_gyration} Temperature dependence of the ensemble-averaged radius of gyration of ring-polymers $r_\t{G}$ for a perfect bcc lattice of $\alpha$-iron. $r_\t{G}^\t{free}$ is the radius of gyration of a free particle of the same mass. The radius of gyration is a measure of the quantum dispersion of the atomic wave function.}
    \end{figure}

    As was previously observed in the context of quantum diffusion\cite{inverse_isotope_effect} and glass transition in quantum liquids\cite{rpmd_glass_1,rpmd_glass_2}, quantum dispersion can hinder atom re-arrangements because in condensed-phase systems the wave-packet of an atom is contracted due to its interaction with neighbor atoms. In other words, each atom is confined/squeezed by its neighboring atoms which reduces atom dimensions below its free-particle size. In RPMD, how much confinement from its neighbors an atom sees can be assessed by computing the average radius of gyration $r_\t{G}$ of the ring-polymers representing the atoms. Shown in the inset of fig.~\ref{fig:radius_of_gyration} is a plot of $r_\t{G}$ in a perfect crystal computed for the same model of $\alpha$-iron as a function of temperature (see Methods \ref{sec:methods_zpe_rg}). That atomic confinement is significant can be gauged from another plot in the same figure showing the ratio of $r_\t{G}$ to the radius of gyration of a free atom $r_\t{G}^\t{free}=\sqrt{\hbar^{2}/m\kB T}/2$ as a function of temperature. When it comes to thermally-activated mechanisms that require overcoming an energy barrier, such as dislocation motion, the already squeezed atoms have to additionally contract their wave packets as they pass through narrow paths leading over the barrier, this additional contraction resulting in an increase in both kinetic and potential energy of atomic configurations corresponding to the barrier states. In the temperature range between $5\K$ and $50\K$ explored in our RPMD simulations, $r_\t{G}$ increases by $30\%$ with decreasing temperature partially offsetting the $50\%$ increase in the energy of zero-point vibrations over the same temperature interval (see fig.~\ref{fig:energy_vs_temperature}).

    As a method for computing equilibrium statistical properties of quantum systems, such as the ZPE and $r_\t{G}$, RPMD is asymptotically exact in the limit of continuous Feynman paths ($P\rightarrow\infty$). On the other hand, motion dynamics of ring-polymers in RPMD is fictitious and does not necessarily reproduce dynamics of quantum motion of atoms. Every time RPMD is used to study dynamic properties, such as dislocation motion of interest here, the validity of simulation results must be carefully examined. Generally, for an RPMD simulation to be a valid representation of quantum dynamics, the simulated RPMD trajectory should consist of discrete transitions from one potential energy well (basin) to another. Further, the system should reside for a sufficiently long time within each energy basin before transitioning to the next basin, to forget about its preceding dynamic trajectory. This is precisely as expected in the equilibrium quantum TST for which RPMD was recently proven to be exact\cite{qtst,rpmd_qtst_3,rpmd_qtst_1,rpmd_qtst_2,rpmd_theory_2,phdthesis}. For the classical TST to be applicable, the residence time within each basin should exceed the thermalization time. Additionally, when atomic motion is quantum, the same residence time should be longer than the quantum decoherence time (see Supplementary Information 3). In our RPMD simulations, a screw dislocation moves in a stop-and-go fashion at the prescribed average velocity of $50\,\t{m/s}$, as illustrated in fig.~S5 (see Methods \ref{sec:methods_dislocation_motion}). Given the distance between two neighboring energy basins (Peierls valleys) of $2.5\Ang$ in $\alpha$-iron, this velocity dictates that the dislocation spends on average $5\,\t{ps}$ in each Peierls valley, including the residence and the transition segments of its RPMD trajectory as shown in fig.~S6. The quantum thermal time $\beta\hbar$ becomes longer than $5\,\t{ps}$ below $T=5\K$ defining the lower bound temperature of our RPMD simulations. The upper bound of $50\K$ is chosen judiciously, since we observed that at temperatures above $50\K$ most of the $5\,\t{ps}$ time is spent in transition between the Peierls valleys rather than in residence in one of them. To extend the trust range of our RPMD simulations to temperatures above $50\K$, the dislocation should be allowed to spend longer time in each Peierls valley necessitating simulations at still lower dislocation velocities.

    Our RPMD simulations of quantum motion of dislocations are unprecedented in scale and accuracy. Indeed, all previous RPMD simulations dealt with systems not exceeding a few thousand atoms at temperatures typically above $0.1$ of the Debye temperature $T_\t{D}$. By comparison, our RPMD simulations entail extended crystal defects embedded in a crystal lattice comprised of $\approx\!150,000$ atoms at temperatures as low as $0.01 \, T_\t{D}$. But RPMD method's rigor comes at a great computational cost: to achieve convergence of quantum RPMD simulations under such conditions, up to $200$ replicas of the entire $150,000$ atom system had to be integrated using time steps comprising a small fraction of one femtosecond over many nanoseconds. In terms of its raw computational cost, each RPMD simulation takes hundreds and thousands times more compute cycles per unit of simulated time compared to a classical MD simulation with the same interatomic potential. We achieved the needed computational performance by building our RPMD simulations on top of an existing highly efficient MD code LAMMPS\cite{lammps} and optimizing the resulting RPMD-LAMMPS method for massively parallel computing. To help make the RPMD method more accessible to materials scientists, we make our code freely available and give sufficient technical details of the method's implementation and fundamentals in our Supplementary Information 1 and 2.

    We have developed an efficient implementation of the RPMD method enabling the study of extended defects in materials accounting accurately for quantum corrections to atom dynamics. We employ RPMD for a detailed investigation of the effect of quantum dynamics on dislocation motion in $\alpha$-iron, which had been previously suggested as a source of the notorious discrepancy between calculated and measured values of the Peierls barrier in bcc metals. The present work utilizes RPMD to show that such corrections are smaller than previously suggested, likely due to quantum dispersion partially compensating the effect of ZPE. The results thus establish that the long-standing discrepancy between calculated and experimental estimates of the Peierls stress in bcc metals remains and further work is needed to settle this outstanding issue in physical metallurgy. We speculate that the origin of this discrepancy is not in the intrinsic properties of single dislocations, rather it lies in complexities emerging from the dislocation network\cite{large_MD} that are unavoidable in experimental studies and neglected in most atomistic simulation studies.

%%%%%%%%%%%%%%%%%%%%%%%%%%%%%%%%%%%%%%%%%%%%%%%%%%%%%%%%%%%%%%%%%%%%%%%%%%%%%%%%
% Methods.                                                                     %
%%%%%%%%%%%%%%%%%%%%%%%%%%%%%%%%%%%%%%%%%%%%%%%%%%%%%%%%%%%%%%%%%%%%%%%%%%%%%%%%

  \clearpage
  \begin{center}
    \textbf{\Large METHODS\label{sec:methods}}
  \end{center}

  \section{Molecular Dynamics simulations of dislocation motion\label{sec:methods_MD}}
      All atomistic simulations were performed in a volume shaped as a rectangular prism oriented as shown in the inset in fig.~\ref{fig:stress_vs_T_MD} with dimensions (at $T=0\K$) $L_{[111]}=61.8\Ang$, $L_{[1\bar{1}0]}=145.4\Ang$, and $L_{[11\bar{2}]}=191.2\Ang$ containing $148,500$ atoms. Periodic boundary conditions were employed along the $[111]$ and the $[1\bar{1}0]$ directions, and free-surface boundary conditions were employed along the $[11\bar{2}]$ direction. A $\f{1}{2}\avg{111}$ screw dislocation was inserted in the center of the simulation volume as described in refs.~\citenum{bulatov_book} and \citenum{weicai_PBC}. Atomic positions in the fixed layers at the top and at the bottom of the simulation volume were corrected to enforce translational invariance along the dislocation motion direction ($\uv{y}$-axis or $[1\bar{1}0]$).

      Classical MD simulations were performed using LAMMPS \cite{lammps} code with an embedded-atom method (EAM) potential previously developed for the bcc $\alpha$-phase of elemental iron \cite{gordon_fe}. The simulation timestep was $\Delta t=3\,\t{fs}$, $\approx 1/40\,\t{th}$ of the period of the highest frequency mode in the phonon spectrum computed for this model system. The Debye temperature, $T_\t{D} = (400\pm20)\K$, was estimated from the same phonon spectrum, as described in ref.~\citenum{debye_temp}. Temperature was maintained near-constant (NVT ensemble) in each simulation using a Langevin thermostat with the temperature relaxation time of $30\,\t{ps}$, which was found to be sufficiently gentle not to affect the flow stress when compared to corresponding microcanonical simulations. Classical NVT simulations were performed at temperatures ranging from $0.1\K$ to $300\K$ using per-atom volumes adjusted at each temperature to maintain internal pressure close to zero. Prior to running mobility simulations, the model was thermalized at the desired temperatures for $300\,\t{ps}$ after which two rigid layers (slabs) of atoms at the top and the bottom of the simulation volume (blue atoms in fig.~\ref{fig:stress_vs_T_MD}) were displaced at a constant velocity in the opposite directions along the $\uv{x}$-axis ($[111]$). The slab velocity was adjusted so as to make the dislocation move at a certain average velocity ($50\,\t{m/s}$ or $10\,\t{m/s}$) once its motion becomes steady after an initial transient period. The straining rate was kept constant for $10\,\t{ns}$ over which the $\tau_{xz}$ component of the internal stress was recorded as a measure of the flow stress. A stress versus time curve obtained in one such MD simulation is shown in fig.~S7.

    \section{Ring-Polymer Molecular Dynamics simulations\label{sec:methods_RPMD}}
      The RPMD simulations were performed using our own implementation of the method in LAMMPS (Supplementary Information 1 and 2). Except for the significantly greater number of degrees of freedom -- $3N$ in MD and $3NP$ in RPMD -- and a shorter time step, all other details of RPMD simulations were the same as in the classical MD simulations described in the preceding section (Methods \ref{sec:methods_MD}).

      The number of beads $P$ (system's replicas) was selected individually for each temperature based on convergence of the system's energy and internal stress-tensor with the number of beads. Figure S8 presents data obtained in one such convergence study at $T = 30\K$. Based on this data we elected to use $40$ beads for RPMD simulations at this particular temperature. Based on similar analyses we ended up using $200$, $100$, $50$, $40$, $30$, and $20$ beads at temperatures $5\K$, $10\K$, $20\K$, $30\K$, $40\K$, and $50\K$, respectively. The timestep was set equal to $1/25$th of the period of the highest frequency eigenmode of the ring-polymers \cite{tuckerman}, which in practice resulted in time steps of $0.48\,\t{fs}$ for $5\K$, $10\K$, $20\K$, and $30\K$, and $0.40\,\t{fs}$ for $40\K$ and $50\K$. Similar to the classical MD, we adjusted the per-atom volume at each temperature in the quantum system to maintain its internal pressure close to zero; in practice the difference in the zero-pressure lattice parameter between MD and RPMD was only of about $\pm0.05\%$ (at the same temperature) and had no appreciable effect on dislocation motion.

      To induce dislocation motion and to compute the Peierls stress we used the same procedures as in the classical MD simulations. To additionally verify that our RPMD predictions are accurate, we repeated RPMD simulations at $T=40\K$ and $50\K$ using twice as many beads as was judged necessary for fully converged simulations and observed no appreciable changes in the results. Additionally, we verified that our flow stress predictions are converged with respect to time step by repeating our RPMD simulations at $T=30\K$ with time steps of $0.20$ and $0.10\,\t{fs}$, i.e. $\frac{1}{2}$ and $\frac{1}{4}$ of the $0.40\,\t{fs}$ deemed acceptable at this temperature.

    \section{RPMD calculations of the zero-point energy and the radius of gyration\label{sec:methods_zpe_rg}}
      RPMD simulations of a perfect bcc lattice of iron with $5,488$ atoms were used to compute the ZPE and $r_\t{G}$ shown in figs.~\ref{fig:energy_vs_temperature} and \ref{fig:radius_of_gyration}. Simulations were performed at the same temperatures and with the same numbers of beads $P$ as in Methods \ref{sec:methods_RPMD}. Additionally, RPMD simulations were performed at higher temperatures: $100\K$ (with 16 beads), $150\K$ (12 beads), $200\K$ ($12$ beads), $250\K$ ($10$ beads), $300\K$ ($9$ beads), $400\K$ ($8$ beads), $500\K$ ($7$ beads), and $600\K$ ($7$ beads). The classical energies shown on the plot were extracted from MD simulations of the same system at the same temperatures. The radius of gyration $r_\t{G}$ of the ring-polymers in RPMD shown in fig.~\ref{fig:radius_of_gyration} was computed as
      \begin{equation}
        r_\t{G}=\Avg{\f{1}{P}\sum_{i=1}^{P}\abs{\v{r}_{n,i}-\bar{\v{r}}_{n}}^{2}}^{1/2}
      \end{equation}
      where $\avg{\ldots}$ denotes the canonical ensemble average and
      \begin{equation}
        \bar{\v{r}}_{n}=\f{1}{P}\sum_{i=1}^{P}\v{r}_{n,i}\label{eq:methods_centroid}
      \end{equation}
      is the ring-polymer centroid. The radius of gyration of the free atom shown in the same figure was computed analytically\cite{rpmd_hydrogen}:
      \begin{equation}
        r_\t{G}^\t{free}=\f{1}{2}\sqrt{\f{\hbar^{2}}{m\kB T}}.
      \end{equation}

    \section{Dislocation motion analysis\label{sec:methods_dislocation_motion}}
      For RPMD simulations to be representative of stochastic dynamics of dislocation motion, it was necessary to assess whether the dislocation spends sufficient time in its energy basins between basin-to-basin transitions. The analysis was performed on the snapshots of atom centroid positions, eq.~\eqref{eq:methods_centroid}, saved every $1.4\,\t{fs}$ along the simulated RPMD trajectories. Within each time snapshot the dislocation's position was taken as a center of mass of atom centroids that were deemed ``defective'' according to the adaptive Common Neighbor Analysis\cite{cna,ovito}; in our low-temperature RPMD simulations most such ``defective'' atoms belong to the dislocation core. Figure S6a is an example of a dislocation trajectory extracted from one of our RPMD simulations. The trajectory reveals a stop-and-go character of dislocation motion in which periods of residence in the energy basins (horizontal segments) are interrupted by more or less fast transitions between the basins (step segments). Duration of the residence periods decreases with increasing temperature (fig.~S6b) and at temperatures above $50\K$ distinct residence periods are no longer recognized. Short of running RPMD simulations at a lower dislocation velocity (which would extend the average residence periods but also raise the computational cost) we no longer consider our RPMD simulations at temperatures above $50\K$ to be representative of dislocation motion in our model of $\alpha$-iron.

  \section*{Acknowledgments}
    The authors would like to thank Alfredo Correa for helpful suggestions concerning calculations of the Debye temperature. 

  \section*{Competing interests}
    The authors declare no competing financial or non-financial interests.

  \section*{Authors contributions}
    VVB designed and planned the project. RF implemented and tested the RPMD method in LAMMPS and performed the simulations. RF, MA, and VVB analyzed simulation results and wrote the paper.

  \section*{Funding}
    RF was supported by the Livermore Graduate Scholar Program. MA was supported by the U.S. Department of Energy, Office of Science, Office of Basic Energy Sciences, Materials Sciences and Engineering Division under Contract No. DE-AC02-05-CH11231 (Mechanical Behavior of Materials Program (KC13)). VVB was supported by the ASC PEM Program at the Lawrence Livermore National Laboratory. Computing support came from the Lawrence Livermore National Laboratory Institutional Computing Grand Challenge program. This work was performed under the auspices of the US Department of Energy by Lawrence Livermore National Laboratory under contract W-7405-Eng-48.

  \section*{Data availability}
    The code used to generate the results in this paper can be downloaded from \url{https://github.com/freitas-rodrigo/RPMDforLAMMPS}. All other relevant data can be obtained directly from the authors.

  \bibliography{bibliography}
\end{document}